\documentclass[sigconf, 10pt, nonacm]{acmart}
\usepackage{microtype}
\usepackage{amsmath}

\usepackage{tikz,array}
\usetikzlibrary{matrix, shapes.geometric, arrows, calc, fit, positioning, chains, arrows.meta, calc, spy, shapes.geometric, arrows, external, patterns}
\usepackage{pgfplots}
\usepackage[caption=false]{subfig}

\usepackage{graphicx}

\pgfplotsset{compat=1.18}
\usepackage{xspace}
\usepackage{booktabs}
\usepackage{ifthen}
\usepackage{multirow}
\usepackage{siunitx}
\usepackage{hyperref}
\usepackage[capitalise,nameinlink,noabbrev]{cleveref}
\crefname{section}{\S{}}{\S{}}
\crefname{figure}{Fig.}{Fig.}
\crefname{table}{Tab.}{Tab.}
\crefname{appendix}{App.}{App.}
\crefname{equation}{Eq.}{Eq.}
\crefformat{section}{\S#2#1#3}
\usepackage{siunitx}
\DeclareSIUnit{\bit}{b}
\DeclareSIUnit{\gbps}{\giga\bit\per\second}
\DeclareSIUnit{\byte}{B}
\usepackage[inline]{enumitem}
\usepackage[dvipsnames]{xcolor}
\usepackage[many]{tcolorbox}
\newtcolorbox{mybox}{
    enhanced,
    sharp corners,
    breakable,  %
    borderline west={1pt}{0pt}{black!70},
    borderline north={1pt}{0pt}{black!70},
    borderline east={1pt}{0pt}{black!70},
    borderline south={1pt}{0pt}{black!70},
    colback=black!10,  %
    colframe=black!10,  %
    before skip=1mm, after skip=1mm,
    boxsep=-1mm
}

\newcommand{\allreduce}[0]{AllReduce\xspace}

\newcommand{\alltoallv}[0]{AllToAll-V\xspace}

\newcommand{\directconnect}[0]{direct-connect\xspace}

\usepackage[suppress]{color-edits}

\addauthor[Mark]{ms}{red}
\addauthor[Ori]{oc}{blue}
\addauthor[Daniel]{da}{Bittersweet}
\addauthor[Jakob]{jk}{purple}

\newcommand{\projname}[0]{{MoX}\xspace}
\title{{\projname:} Efficient MoE Routing on Direct-Connect Topologies}

\author{Ori Cohen}
\affiliation{%
  \institution{Technion and NVIDIA}
  \country{Israel}
}
\email{ori.cohen@campus.technion.ac.il}

\author{Jakob Krebs}
\affiliation{%
  \institution{Technion}
  \country{Israel}
}
\email{jakob.krebs@campus.technion.ac.il}

\author{Daniel Amir}
\affiliation{%
  \institution{Technion}
  \country{Israel}
}
\email{daniel.amir@campus.technion.ac.il}

\author{Mark Silberstein}
\affiliation{%
  \institution{Technion and NVIDIA}
  \country{Israel}
}
\email{mark@ee.technion.ac.il}

\renewcommand{\shortauthors}{Cohen et al.}

\begin{abstract}
Optically switched networks suit the regular communication of dense ML models,
but MoE introduces sparse, runtime-dependent traffic. We show that efficient offline-optimized routing enables efficient MoE training and inference on direct-connect topologies
without the need for MoE traffic matrix or dynamic topology reconfiguration. \projname constructs token-aware
multicast trees to reduce bandwidth tax, then uses static, precomputed
link weights to balance traffic by solving a restricted multicast tree-packing problem. Using recorded traffic from large MoE models,
token-level traces, and ASTRA-sim, we find that \projname accelerates the full
MoE block---dispatch, expert computation, and combine---by up to $1.8\times$ over min-hop routing. Moreover,  it attains nearly ideal packet-switched network performance in random expander topologies. On a 1,024-TPU model of Google's Boardfly
topology, \projname reduces the dispatch bottleneck link load by up
to $47\%$. These results show that high-performance MoE on static direct-connect fabrics can be achieved via optimized load-oblivious routing without demand-driven reconfiguration.
\end{abstract}

\begin{document}\sloppy
\raggedbottom
\maketitle
\section{Introduction}
The performance of large machine-learning systems increasingly depends on their networks.
High-bandwidth domains (HBDs) such as NVLink provide abundant bandwidth, but are expensive and difficult to scale. Once a job extends beyond the HBD, communication over the scale-out network often becomes a bottleneck.

Optical networks offer a way to expand HBDs with significantly lower power and cost than a conventional
packet switch at every tier. Direct-connect and optical circuit-switched (OCS) networks 
are already used inside production training and inference systems~\cite{tpuv4,tpu8i}, and recent
proposals demonstrate the benefits of reconfiguring the scale-out fabric for individual
collectives~\cite{photonic-rails} or interconnecting multiple smaller
HBDs~\cite{mixnet}.
Because optical networks directly connect nodes to each other, they are particularly attractive when the communication pattern is static and regular, as is the case for data-, tensor-, and pipeline-parallel training.
A \directconnect fabric can be structured to match this stable collective structure, providing strong performance at production scale in industry today~\cite{scaling-transformer-inference,inferentia2,groq,tpuv4,tpu8i}.

Mixture-of-Experts (MoE) models break this regularity. In every MoE layer, a
gate selects, independently for each token, its top $K$ experts out of $E$.
When experts are distributed across accelerators, the resulting
expert-parallel (EP) communications are sparse \alltoallv: each accelerator
dispatches tokens to a data-dependent subset of peers and later collects the
expert results.
The demand changes with every batch, can span the full EP
group, and is skewed because some experts receive more tokens than others.
Unlike a ring or an \allreduce, this traffic does not lend itself to one stable,
structured topology.

Expert domains are rapidly outgrowing commodity HBDs: Mixtral has eight
experts per layer~\cite{mixtral}, Qwen3 and LLaMA~4 have 128~\cite{qwen3,llama4},
DeepSeek-V3 and Pangu use 256~\cite{deepseekv3,pangu}, Qwen3.5 reaches
512~\cite{qwen35}, and Kimi~K3 uses 896~\cite{kimik3}. The number of active experts (top-k) is also increasing, with DeepSeek-V3 using $k=8$, Qwen3.5 using $k=10$, and Kimi~K3 using $k=16$~\cite{deepseekv3,qwen35,kimik3}, as are EP group sizes, with DeepSeek-V3 already being trained using 64-way EP~\cite{deepseekv3,qwen3}.
A low-degree direct-connect fabric cannot provide single-hop connectivity to all
such destinations.
At the same time, OCSes reconfigure too slowly to be operated during a collective~\cite{mixnet,photonic-rails}.
While MixNet~\cite{mixnet} has combined runtime-optimized direct optical links with packet switches for traffic lacking a direct circuit, this strategy greatly limits the utility of the optical network when system scale exceeds the degree at each node.
Efficient optical MoE therefore requires \emph{indirect routing}, even with topology specialization.

Indirect routing creates two problems. First, every relay hop consumes bandwidth
without delivering data locally, resulting in \emph{bandwidth tax}. Second, relay traffic
may be distributed unevenly across links, resulting in \emph{load imbalance}. This occurs
in irregular topologies such as random expanders and in nominally regular
topologies whose physical connectivity is asymmetric, such as TPU~8i Boardfly.
Skewed MoE demand further amplifies these structural hotspots. Their effect is
significant because collective completion time is determined by the most loaded
link: in our evaluation on a DeepSeek-V3 MoE traffic, the busiest link carries up to $1.35\times$ the average
link load within Boardfly's network.

These two problems expose a topology tradeoff. Common structured topologies, such
as 3D-torus, use symmetry to distribute relay traffic evenly, but at fixed
node degree their relatively high radius requires more relay hops and therefore
a larger bandwidth tax. Low-radius irregular topologies remedy the first
problem. Random expanders connect distant endpoints in few hops at low degree
and can scale to large systems~\cite{amazon-expander}; however, their irregular
paths make some links more likely to relay traffic, worsening the second
problem. We use expanders to develop this intuition, but our techniques
apply to any topology. 

We present \emph{\projname}\footnote{Pronounced ``Mo-X''.}, a static, demand-oblivious MoE routing system that uses
EP semantics and precomputed top-K-specific per-link weights instead of predicting traffic matrices or
reconfiguring the network. Its two techniques address indirect routing's
costs:

\noindent\textbf{Token-aware multicast and reduction trees.}
MoE dispatch is a collection of per-token
multicasts: the same token must reach the $K$ accelerators hosting its
selected experts. \projname performs indirect routing of a token via nodes which also receive that token, 
forming a multicast tree.  During combine, it reverses these trees and partially reduces
expert outputs at intermediate accelerators. Similar techniques were applied in prior work 
on efficient collectives in \directconnect topologies \cite{bier,xcast}, but \projname is the first to  
show that applying them to MoE in this way is highly effective for reducing the bandwidth tax.

\noindent\textbf{Precomputed skew-oblivious load-balanced routing.}
\projname formulates relay selection as a compact surrogate for the fractional
multicast-tree packing problem studied in traffic
engineering~\cite{multicast-congestion,multicast-packing}. Each source and
top-$K$ expert selection defines a multicast demand class; the selected experts
are mapped to their hosting accelerators, inducing a set of candidate multicast
trees. Rather than predict the deployed traffic matrix, \projname assigns equal
likelihood to all top-$K$ expert combinations and computes link routing weights, one weight per link,
that minimize the maximum expected physical-link load over the resulting tree
distribution.  
Enumerating all destination
sets and trees is combinatorial, so
\projname computes the weights offline from a small sample of multicast
workloads. The resulting policy remains effective on real, non-uniform MoE
traffic for the given choice of $K$, without runtime demand prediction or per-iteration topology adaptation.

Expert popularity is often skewed~\cite{lina,mixnet}, and both load and its
variation differ across MoE layers~\cite{moe-load-stability}. Nevertheless,
modern MoE training encourages balanced marginal utilization
through auxiliary losses~\cite{switch-transformer} or load-dependent routing
biases~\cite{wang2024auxiliary,deepseekv3}. Moreover, experts from different
layers are randomly mapped to accelerators, so a hot expert in one layer
need not create a hotspot at the same physical endpoint in another. Thus, after
aggregating layers under a randomized mapping, we expect a uniform prior to
provide a useful topology-level center for offline optimization.

We evaluate \projname using token-level Chakra traces in
ASTRA-sim~2~\cite{chakra,astrasim2},  with DeepSeek-V3 and Qwen-3 MoE layers and traffic
distributions recorded on a real GPU cluster.

We make three observations:

\noindent {\bf Near-switch performance.} For recorded traces, on 16-, 32-, and 64-node degree-8
expanders, top-12 training is within 0.5\%, 0.6\%, and 7\% of an ideal switch,
while min-hop routing is 42\%, 69\%, and 96\% slower. With 256 inference tokens
per GPU, \projname is within 2.1\%, 3.1\%, and 13.6\%, while min-hop is
42--98\% slower.

\noindent {\bf Better than topology adaptation.} On recorded traffic, static
\projname routing achieves 27\% lower completion time than a demand-optimized
topology with min-hop routing and nearly attains the perfect switch performance across evaluated skews.

\noindent {\bf Boardfly load balancing.} On a regular
1,024-TPU hierarchy constructed from Google's published TPU~8i Boardfly
parameters, where connectivity constraints introduce endpoint-level
asymmetries, \projname reduces bottleneck cable load by up to 47\%.

\noindent\textbf{Bottom line.} \projname shows that high-performance MoE communication is achievable without continuously adapting the physical network to runtime demand: a static
direct-connect topology can approach switched-network performance when routing
is designed around MoE's multicast semantics and load-balanced offline.

\section{Routing Algorithm}

\begin{figure}
    \centering
    \includegraphics[width=\linewidth]{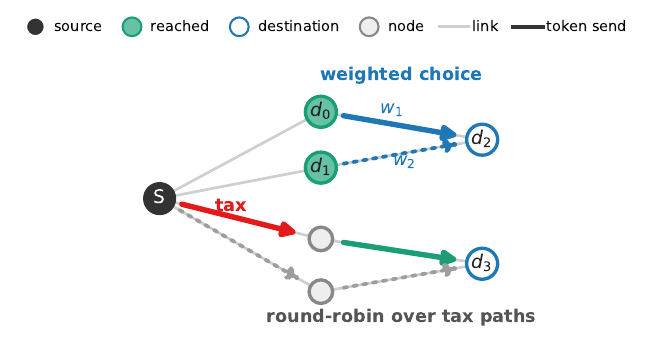}
    \caption{Routing an embedding whose expert destinations are
    $d_0$--$d_3$. Once $d_0$ and $d_1$ hold the embedding, either can relay it
    to adjacent destination $d_2$; \projname selects between them using link
    weights. No reached destination neighbors $d_3$, so reaching it requires a
    non-destination relay and incurs a bandwidth-tax hop.}
    \label{fig:token-routing}
\end{figure}

\projname routes MoE token embeddings over a fixed, known network topology.
For every token, it first constructs a multicast tree that minimizes redundant
forwarding and then selects among equivalent relays using offline-computed link
weights. During combine, it reverses the tree and partially reduces expert
outputs at intermediate nodes. 
\Cref{fig:token-routing} shows a simple example of the routing process.
Throughout this section, dispatching a
\emph{token} means transmitting its MoE-layer embedding (activation vector).

\subsection{Building the Multicast Tree}
\label{sec:token-forward}

For a token generated at source~$s$, let $D$ denote the accelerators hosting
its top-$K$ experts. \projname constructs a directed multicast tree rooted
at~$s$ and spanning~$D$. It maintains a holder set~$H$, containing the nodes
that already have the embedding, and initializes $H=\{s\}$.

At each step, \projname selects an unreached destination $d\in D\setminus H$
and a path from a node in~$H$ to~$d$. It prefers a path whose intermediate
nodes are themselves members of~$D$: those nodes require the embedding and can
subsequently relay it to another destination. More generally, it selects paths
that minimize the number of intermediate nodes outside~$D$. The selected path
is added to the multicast tree and its nodes are added to~$H$. This process
continues until all destinations have been reached.

The two-hop case illustrates the rule. If a selected expert~$r$ lies on a path
from~$s$ to another selected expert~$d$, \projname sends one copy along
$s\rightarrow r\rightarrow d$. Because $r$ consumes the embedding as well as
forwarding it, both hops perform useful delivery. If no such destination relay
exists, \projname uses a non-destination relay and incurs one bandwidth-tax
hop. Ties among paths with the same tax are resolved by the load-balancing
policy in \cref{sec:lp-routing}.

The source controls the tree and encodes the remaining destinations as a
bitmap in the packet header, similarly to stateless multicast mechanisms such
as Xcast~\cite{xcast} and BIER~\cite{bier}. A relay forwards one copy along
each outgoing tree edge. At 64 endpoints, the bitmap occupies 8~bytes, compared
with a 14\,KB token embedding.

\subsection{Load Balancing with Per-Link Weights}
\label{sec:lp-routing}

Tree construction often exposes several relay choices with the same bandwidth
tax. Consider an unreached destination~$d$ and let
$C=H\cap N(d)$ be the current holders adjacent to~$d$. Every $r\in C$ can
deliver the embedding to~$d$ in one hop. \projname associates a positive
weight $w_{(r,d)}$ with every directed physical link and assigns the relay
share
\begin{equation}
    p(r\mid C,d)
    = \frac{w_{(r,d)}}{\sum_{r'\in C}w_{(r',d)}}.
    \label{eq:relay-prob}
\end{equation}
The weights are shared across all tokens and candidate sets; \projname does not
store a separate policy for each~$C$.

At runtime, \projname realizes these shares with weighted round robin (WRR).
Each source maintains per-link counters for the duration of one MoE collective.
For candidate set~$C$, it selects the relay minimizing
\begin{equation}
    \frac{\operatorname{counter}(r,d)}{w_{(r,d)}}
    \qquad r\in C,
    \label{eq:wrr}
\end{equation}
and then increments the selected link's counter. The counters are reset at the
start of the next collective. Over many tokens, this deterministic policy
approaches the shares in \eqref{eq:relay-prob}. A single weight per directed
link keeps the policy state proportional to the physical topology; for example,
a 64-node degree-8 topology requires 512 weights.

When no destination relay is available, \projname selects among equal-tax
first hops using round robin. We treat the load produced by these
unavoidable paths as background load when computing the WRR weights. Paths
with multiple non-destination relays are handled analogously.

\subsection{Fractional Multicast-Tree Packing}

The ideal load-balancing problem is a fractional multicast-tree packing
problem~\cite{multicast-packing,multicast-congestion}. Let $Q$ denote the set
of multicast demand classes. A class $q\in Q$ specifies a source, a top-$K$
destination set, and expected demand~$a_q$. Let $\mathcal{T}_q$ be the set of
minimum-tax multicast trees admitted by the forwarding algorithm for~$q$.
For each $T\in\mathcal{T}_q$, variable $x_{qT}\geq0$ is the fraction of
class-$q$ demand routed on tree~$T$. With directed-link capacity~$c_e$, the
minimum-congestion packing is the linear program
\begingroup
\setlength{\abovedisplayskip}{3pt}\setlength{\belowdisplayskip}{3pt}\setlength{\jot}{1pt}
\begin{align}
    \min_{x,L_{\max}}\quad & L_{\max} \label{eq:mtp-obj} \\
    \text{s.t.}\quad
    & \sum_{T\in\mathcal{T}_q}x_{qT}=a_q
        && \forall q\in Q, \label{eq:mtp-demand}\\
    & \sum_{q\in Q}\sum_{T\in\mathcal{T}_q:e\in T}x_{qT}
        \leq c_e L_{\max}
        && \forall e\in E. \label{eq:mtp-cap}
\end{align}
\endgroup
The first constraint routes all expected demand; the second bounds every
physical link's utilization. The objective minimizes the utilization of the
busiest link. This formulation is closely related to classical multicast
packing and minimum multicast-congestion formulations, which likewise assign
concurrent multicast demands to Steiner trees while minimizing their maximum
edge congestion~\cite{multicast-packing,multicast-congestion}.

The LP is not practical to materialize: $\mathcal{T}_q$ is exponential, and
the number of top-$K$ destination sets is combinatorial. It would also require
candidate-set- or tree-specific routing state at runtime. \projname therefore
optimizes a compact surrogate: the tree shares are not independent variables,
but are induced by the per-link weights through \eqref{eq:relay-prob}. The
surrogate retains the LP's min-max link-load objective while reducing the
runtime policy to one value per directed link.

\subsection{Iterative Weight Computation}

\projname computes the weights offline using Monte Carlo workloads. It starts
with uniform weights, $w_e=1$, routes every sampled token using the complete
tree-construction and WRR policy, and measures the resulting directed-link
loads~$L_e$. This simulation includes the fixed round-robin traffic from paths
that require non-destination relays, so the measured load combines useful
forwarding and bandwidth-tax traffic.

After a replay, \projname updates every weight using
\begingroup
\setlength{\abovedisplayskip}{3pt}\setlength{\belowdisplayskip}{3pt}
\begin{equation}
    \log w_e \leftarrow \log w_e
      - \eta\left(\frac{L_e}{\bar L}-1\right),
    \qquad \eta=0.05,
    \label{eq:weight-update}
\end{equation}
\endgroup
where $\bar L$ is the mean directed-link load. An overloaded link loses weight
and is selected less frequently in subsequent replays; an underloaded link
gains weight. The update is a multiplicative min-max controller for the compact
weight policy: it seeks the same low-congestion outcome as
\cref{eq:mtp-obj,eq:mtp-cap}, but does not enumerate or explicitly solve for
the LP's tree variables. We retain the weight vector from the replay with the
lowest observed maximum link load.

\paragraph{Sampling configuration.}
\projname samples $1{,}000$ tokens per source accelerator; for each, it draws a top-$K$ expert
combination from the assumed uniform workload distribution and applies the
expert-to-accelerator mapping to obtain its destination set. This same sample
is replayed across all six iterations---one with uniform weights followed by
three weight updates using \eqref{eq:weight-update}. We validate that at small sizes with $n=16$, where full
enumeration is feasible, the sampled solution matches the enumerated solution.
In terms of the runtime, at $n=64$, the offline computation completes in seconds on one CPU core.

\subsection{Partial Reduction on Combine}
\label{sec:combine}

After expert computation, the combine phase returns the expert outputs to the
token's source. Semantically, this operation is a reduction: the outputs of the
top-$K$ experts are combined into one activation vector for the next layer.
\projname traverses the dispatch multicast tree in reverse. Whenever an
intermediate node has results from multiple child branches, it partially
reduces them before forwarding one vector toward the source. This avoids
sending every expert output independently across shared links.

\section{Evaluation}
\label{sec:eval}

\begin{figure*}[t]
    \centering
    \subfloat[Training\label{fig:e2e}]{
   
    \includegraphics[width=.45\linewidth]{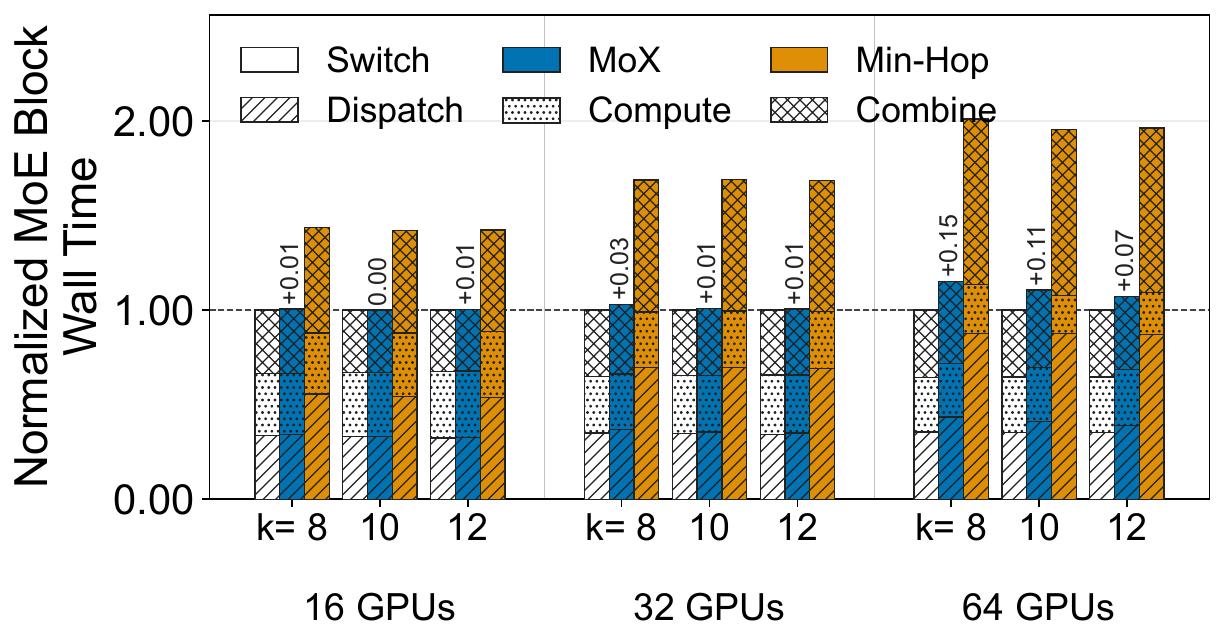}    
}
\hfill
 \subfloat[Inference\label{fig:inference}]{
    \centering
    \includegraphics[width=.45\linewidth]{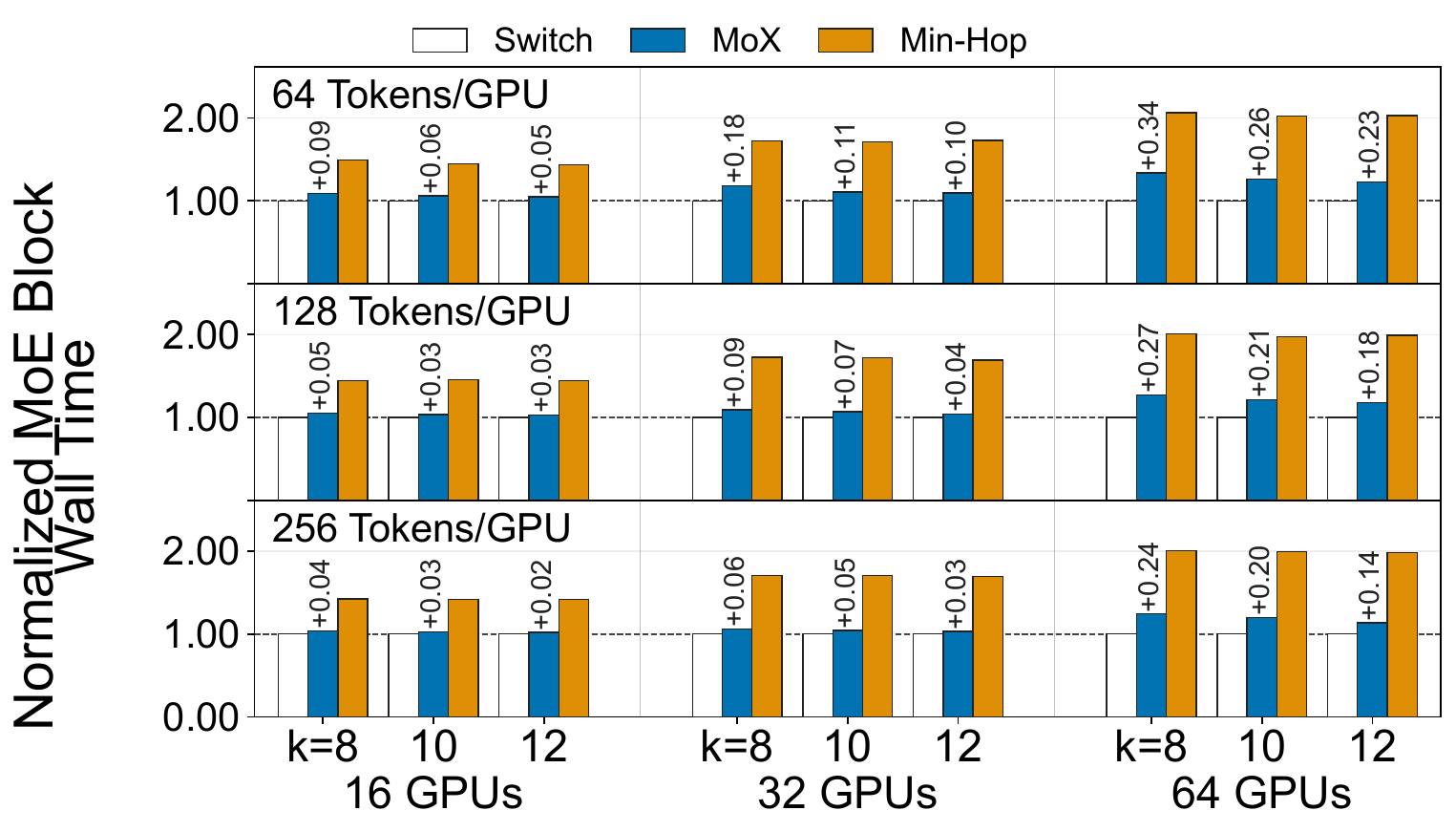}
 }
    \caption{MoE-block wall time normalized to an ideal switch, for the MoE load distribution of DeepSeek-V3.
    \projname has low overhead in 16- and 32-node expanders, and closes the gap at higher Top-k in the 64-node expander.
    }
    \label{fig:e2e-combined}
\end{figure*}

We evaluate \projname with ASTRA-sim~2~\cite{astrasim2} using token-level EP
Chakra traces~\cite{chakra}, representing each routing decision as per-link
transfers. We record all layers in three late-training DeepSeek-V3
runs~\cite{deepseekv3} on 16, 32, and 64 GPUs with 256 experts. In addition, we record the expert popularity for the Qwen-3 225B with 128 experts. Expert popularity is stable over
the sampled iterations and is similar across layers, consistent with prior
observations~\cite{moe-load-stability,mixnet}; we evaluate a randomly selected
layer. In all the experiments the experts are
randomly and evenly placed across GPUs.

Training dispatches 5,120 BF16 embeddings of width 7,168 per GPU; inference
uses the same expert distribution with between 64 and 256 tokens. We compare \projname
and pairwise min-hop routing on bandwidth-equivalent direct-connect fabrics
against an ideal full-connectivity packet switch. Unless noted, each endpoint's
800\,Gbps is split into eight 100\,Gbps links on a randomly generated degree-8
expander selected for low radius. We also evaluate the performance of \projname on the TPUv8i Boardfly topology.

\subsection{MoE end-to-end performance}
\label{sec:eval:collective}

\Cref{fig:e2e-combined} reports full MoE-block
(dispatch--compute--combine) time for DeepSeek MoE. \projname improves over min-hop by up to
$1.8\times$. Training overhead grows with topology size as paths lengthen, but
decreases with higher top-$k$ because more destinations can serve as useful relays.
\projname is within $0.6\%$, $3.0\%$, and $15.1\%$ of an ideal switch for 16, 32, and 64 GPUs at top-$8$, shrinking to $0.5\%$, $0.6\%$, and $7.3\%$ at top-$12$.

Inference follows the same trend but small batches expose fixed transfer costs
and offer less traffic to balance. At 256 tokens per GPU, \projname is within
2.1--3.6\% of the switch on 16 GPUs and 3.1--6.4\% on 32, whereas min-hop is
42--70\% slower (\cref{fig:inference}).

\begin{figure}[t]
    \centering
    \includegraphics[width=\linewidth]{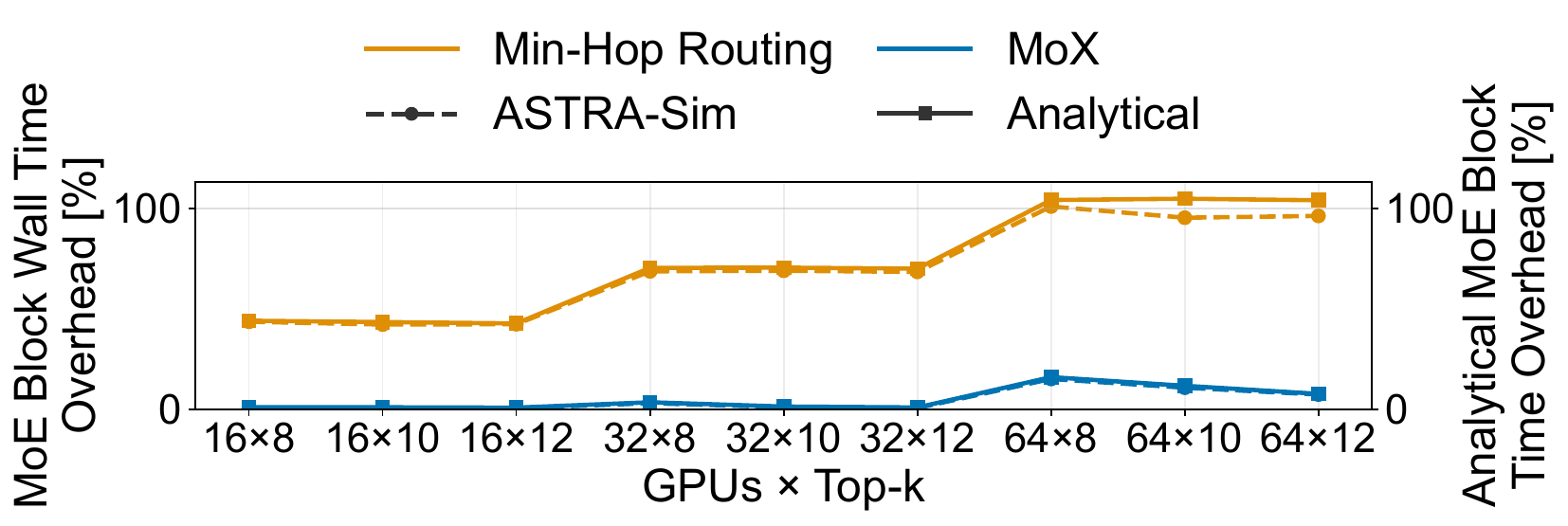}
    \caption{Analytical maximum-link-load overhead and ASTRA-sim end-to-end
    slowdown relative to the switch; the two correlate closely, overlapping in many points.}
    \label{fig:analytical-validation}
\end{figure}

\paragraph{Analytical proxy.} We divide the expander's analytically computed maximum normalized link
load (bytes/capacity) by the switch's one obtained in simulation, and compare it with the ASTRA-sim-reported end-to-end slowdown which simulates the collective on the topology.
The close match in \cref{fig:analytical-validation} validates maximum link load
as a proxy for these bandwidth-bound configurations, which is expected since the maximum loaded link determines the total collective completion time.

\subsection{Sensitivity to load skew and topology optimizations}

\begin{figure}[t]
    \centering
       \includegraphics[width=\linewidth]{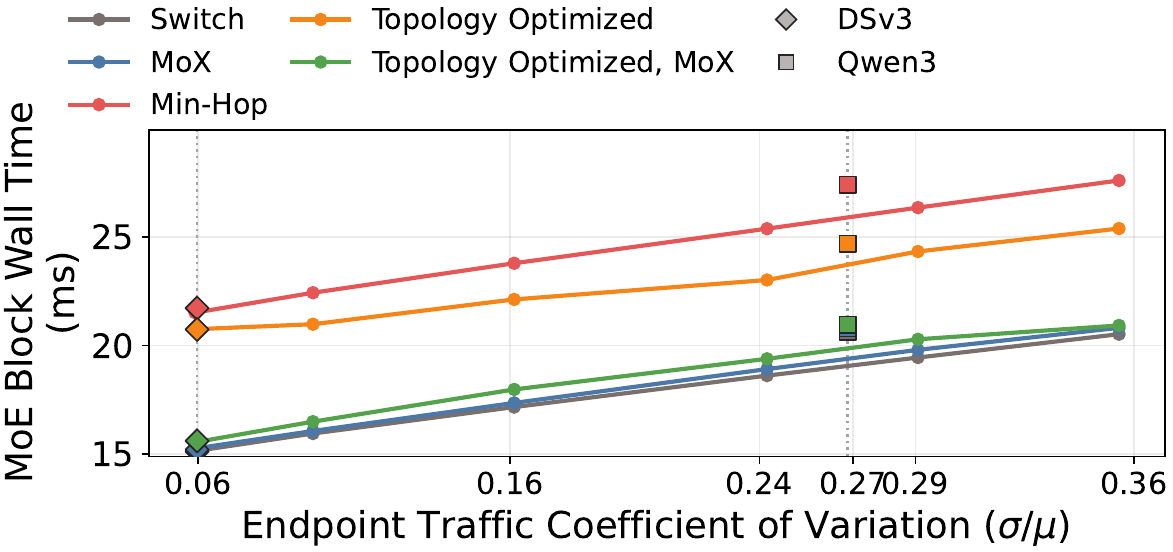}
    \caption{ Performance vs. ideal switch for the synthetic Zipf load distribution, and two real MoE traffic distributions -- DeepSeek-V3 and Qwen 3-225B on a 16-node expander and $K=8$}
    \label{fig:topo-routing:e2e}
    \label{fig:mixnet}
\end{figure}

We compare \projname against the demand-adaptive topology tuning approach inspired by MixNet~\cite{mixnet} on a 16-node expander and top-8 experts.  This approach assumes full knowledge of the traffic matrix: it reconfigures the topology prior to the execution, maximizing the amount of direct traffic thus reducing the bandwidth tax. The tokens are routed via min-hop routing. Note, however, that the number of experts is larger than the node degree, therefore some traffic remains indirect. Last, we combine the topology tuning with the \projname routing approach. 

\Cref{fig:mixnet} presents the results. Across all skews \projname stays within $1.8\%$ of the switch, significantly outperforming both min-hop routing ($34$--$42\%$ slower) and topology optimization ($24$--$37\%$ slower). \projname's routing also rescues the optimized topology, cutting its overhead to $2$--$5\%$; even so, combining the two does not improve on \projname over the plain expander.

\projname's overhead over the switch stays roughly flat across the skew range. The real MoE traffic we recorded for two large-scale models stays well balanced, so \projname remains within $0.8\%$ of the switch.

\subsection{Google TPUv8i Boardfly}
\label{sec:eval:boardfly}

\projname's applicability goes beyond irregular asymmetric topologies.  Here we show that topologies based on more regular graphs are often structured in a way that can create load imbalances when using min-hop routing. As a representative example, Google's \emph{Boardfly} topology used in TPUv8i~\cite{tpu8i} is regular, but doubles the capacity of certain links in order to fully utilize the available bandwidth at each node. Min-hop routing may lead to underutilization of the capacity of these parallel links, resulting in bottlenecks on the remaining links. 
Boardfly uses 400GB/s links~\cite{tpu8-bw}. 

We resample the DeepSeek trace to a projected 4K-expert model, placing four experts on
each of 1K TPUs. Because we do not simulate this full topology end to end, we
report the validated maximum-link-load proxy for dispatch.

\begin{figure}[t]
    \centering
    \includegraphics[width=\linewidth]{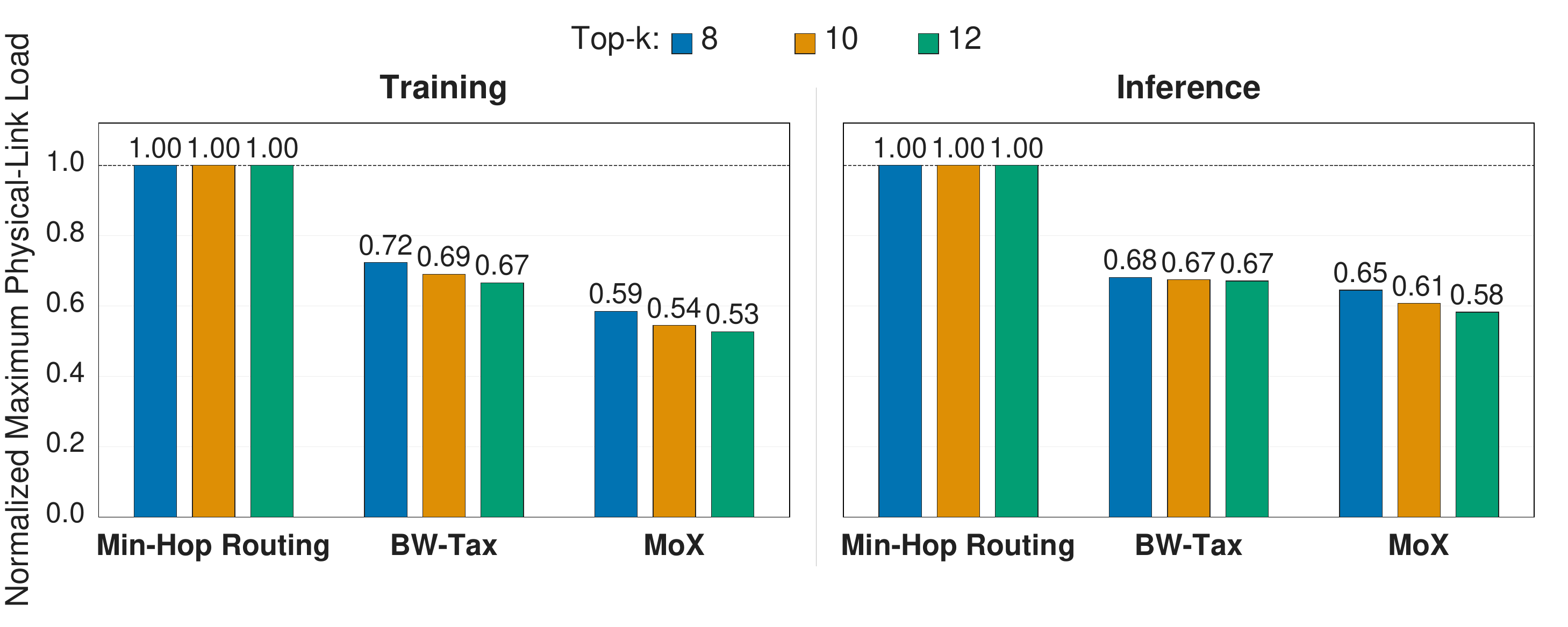}
    \caption{Boardfly dispatch bottleneck: maximum physical-cable load,
    normalized per top-$k$ to shortest-path routing, for training and
    inference. The bottleneck is an inter-group cable in every case.}
    \label{fig:boardfly}
\end{figure}

\Cref{fig:boardfly} shows that \projname reduces the bottleneck link by up to $47\%$ ($33\%$ by relaying tokens and $14\%$ by weighted routing) for training batch sizes and $41.7\%$ ($32.9\%$ by relaying and $9\%$ by weighted routing) for inference batches.

\subsection{Analysis}
\label{sec:eval:scaling}

\begin{figure}
    \centering
    \subfloat[Contribution of routing and bandwidth tax for top-$8$ under different topology sizes
   \label{fig:growing-expander}]{
    \includegraphics[width=.44\linewidth]{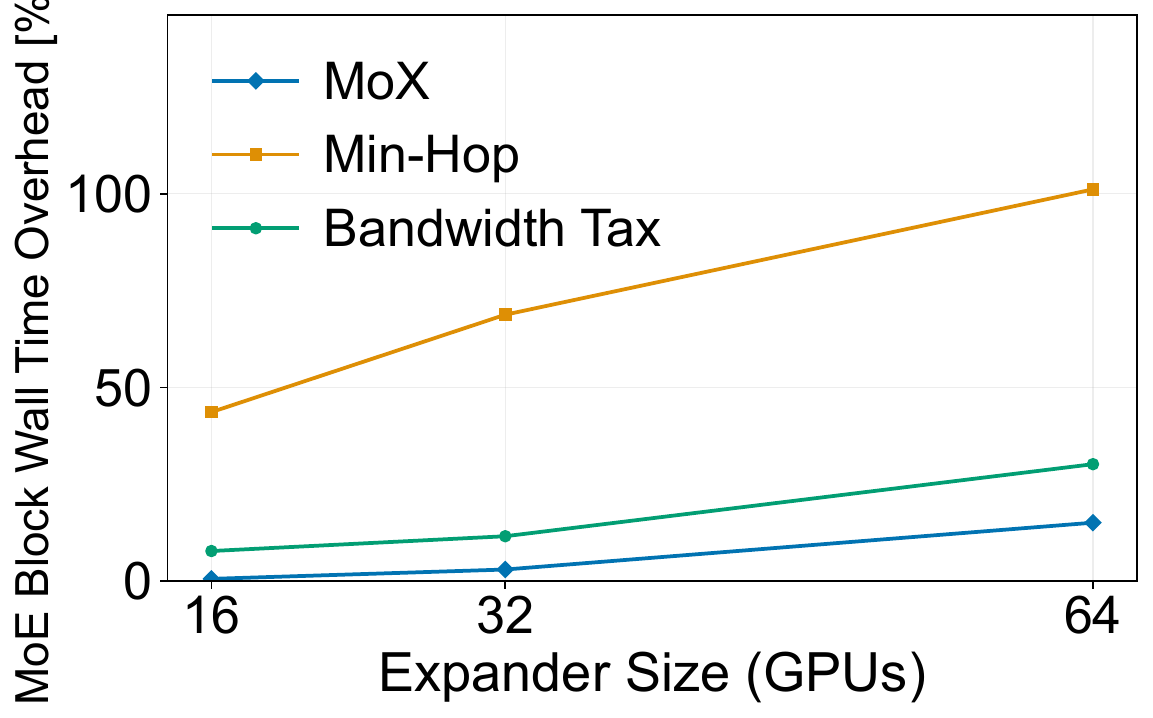}
    }\hfill
    \subfloat[Performance impact of the node degree\label{fig:increase-topk}]{
    \includegraphics[width=.44\linewidth]{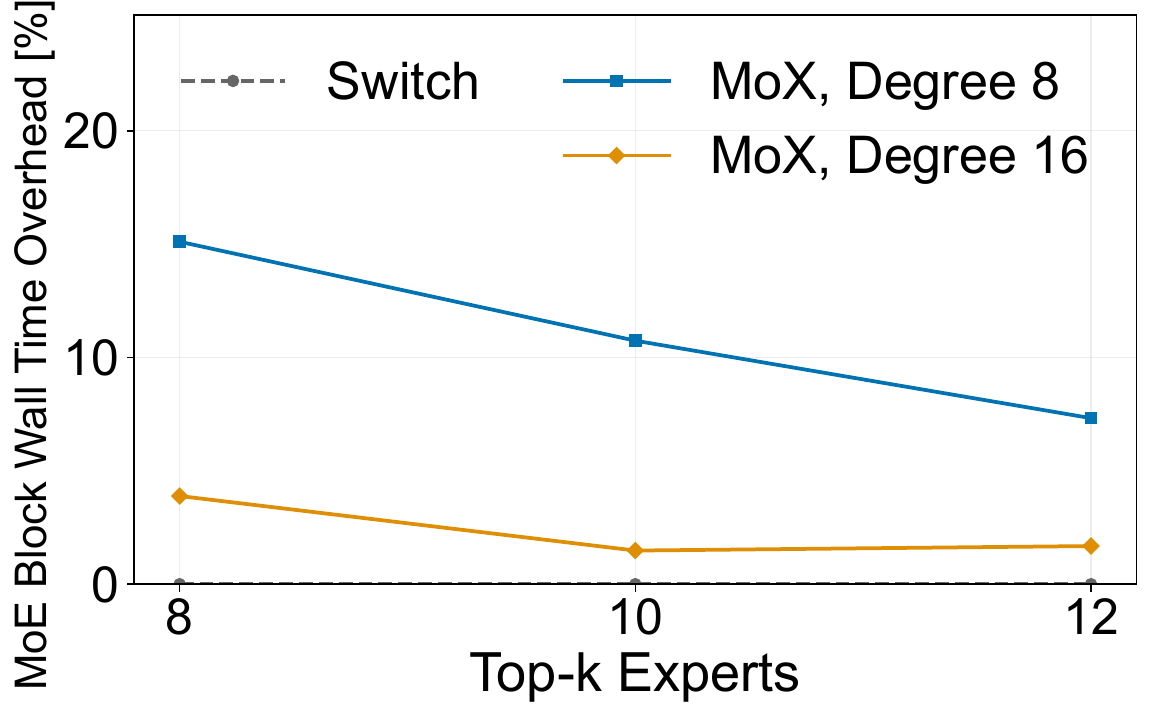}
    }
    \caption{Overhead analysis for different system parameters vs. an ideal switch}
    \label{fig:overhead-breakdown}
\end{figure}

We analyze \projname across expander sizes and top-$k$ values.
\Cref{fig:overhead-breakdown}(a) shows that
the multicast tree alone reduces most of the bandwidth overhead of the min-hop routing down to $8\%$, $11\%$, and $31\%$ for 16, 32, and 64 nodes respectively, but \projname improves further by mitigating the load imbalance. 

With larger topologies, a higher degree expander and higher values of top-$k$ achieve much better results (\Cref{fig:overhead-breakdown}(b)). This suggests that one should maintain degree $d$ that is high enough for the expander topology size $N$, ensuring that the fraction of nodes reachable with a single relay remains constant, e.g. $\frac{d^2}{N} = 2$.

\section{Discussion}

\paragraph{Implementation}
 \projname requires per-token forwarding rather than today's aggregated sends.
We believe that this can be implemented with low overhead due to the large size of each individual token.
Per-token multicast forwarding is accomplished using a destination matrix stored in each packet, allowing it to be handled by either the GPU or NIC.
During combine, reduction must be handled by the GPU, which already stores one of the inputs to the reduction.
Line rate \allreduce has been demonstrated in GPUs~\cite{demystifying-nccl,nccl-allreduce-bench}.

\paragraph{Practicality of high-degree random expander topologies}
When scaling up to high node counts, it is essential for each node to have a high degree as shown in \Cref{fig:growing-expander}.
Cabling has commonly been considered the main barrier to high-degree random topologies.
However, Amazon has recently shown that it has solved this problem in its RNG infrastructure using shuffle boxes~\cite{amazon-expander}.
Further, we expect that the effective number of links per NIC will grow as well, due to increasing adoption of multi-plane topologies~\cite{MRC-network}.

Another approach is to consider each NVLink domain as a single routing endpoint, as suggested in~\cite{mixnet},
using NVLink to share all the NICs in each server. Splitting each of the server's 8 scale-out links into 8 lanes results in a degree of 64, allowing scaling up to 2{,}048 servers (16k total GPUs) while maintaining $\frac{d^2}{N} \ge 2$.

\section{Related Work}

\paragraph{Multicast Tree and Traffic Engineering}
Multicast routing combines NP-hard tree construction
~\cite{steiner-tree-nphard} with traffic engineering that minimizes maximum
link utilization~\cite{micro-te,swan}. Protocols build shortest-path trees
~\cite{pim-sm}, stateless BIER-TE steers static replication paths~\cite{bier},
and controllers adapt routes to group demand~\cite{online-multicast-te}.
MoE changes too quickly for per-iteration weight optimization.

\paragraph{Collectives on Direct-Connect Topologies}
Topology-specific collective algorithms provide near-optimal schedules on
symmetric direct-connect fabrics~\cite{a2a-on-tori,dgx-collectives,
collectives-on-torus,efficient-alltoall-in-direct-connect}, but assume regular,
predictable traffic rather than dynamic, skewed MoE demand.

\paragraph{Direct-Connect Topologies for ML Training}
TopoOpt~\cite{topoopt} co-optimizes topology and parallelization. Google's
optically reconfigurable TPU v4~\cite{tpuv4} targets structured traffic,
whereas TPUv8i adopts the more general Boardfly topology, better suited to
dynamic MoE inference~\cite{tpu8i}. Photonic Rails~\cite{photonic-rails}
time-multiplexes links across training phases; MixNet~\cite{mixnet} predicts
demand to reconfigure a hybrid optical-electrical fabric at runtime.

\section{Conclusion}
\projname demonstrates that routing optimizations which are lightweight, computed offline, and load-oblivious provide a significant performance boost for MoE communication on \directconnect topologies, without the need for costly dynamic topology adoption or continuous adjustment of routing weights. We show that \projname significantly reduces the overheads of the MoE phase for both inference and training, achieving performance close to an ideal packet switch, and demonstrates significant potential for optimizing Google's recent Boardfly direct-connect network.

\bibliographystyle{ACM-Reference-Format}

\begin{thebibliography}{39}


\ifx \showCODEN    \undefined \def \showCODEN     #1{\unskip}     \fi
\ifx \showDOI      \undefined \def \showDOI       #1{#1}\fi
\ifx \showISBNx    \undefined \def \showISBNx     #1{\unskip}     \fi
\ifx \showISBNxiii \undefined \def \showISBNxiii  #1{\unskip}     \fi
\ifx \showISSN     \undefined \def \showISSN      #1{\unskip}     \fi
\ifx \showLCCN     \undefined \def \showLCCN      #1{\unskip}     \fi
\ifx \shownote     \undefined \def \shownote      #1{#1}          \fi
\ifx \showarticletitle \undefined \def \showarticletitle #1{#1}   \fi
\ifx \showURL      \undefined \def \showURL       {\relax}        \fi
\providecommand\bibfield[2]{#2}
\providecommand\bibinfo[2]{#2}
\providecommand\natexlab[1]{#1}
\providecommand\showeprint[2][]{arXiv:#2}

\bibitem[Abts et~al\mbox{.}(2022)]%
        {groq}
\bibfield{author}{\bibinfo{person}{Dennis Abts}, \bibinfo{person}{Garrin
  Kimmell}, \bibinfo{person}{Andrew Ling}, \bibinfo{person}{John Kim},
  \bibinfo{person}{Matt Boyd}, \bibinfo{person}{Andrew Bitar},
  \bibinfo{person}{Sahil Parmar}, \bibinfo{person}{Ibrahim Ahmed},
  \bibinfo{person}{Roberto DiCecco}, \bibinfo{person}{David Han},
  \bibinfo{person}{John Thompson}, \bibinfo{person}{Michael Bye},
  \bibinfo{person}{Jennifer Hwang}, \bibinfo{person}{Jeremy Fowers},
  \bibinfo{person}{Peter Lillian}, \bibinfo{person}{Ashwin Murthy},
  \bibinfo{person}{Elyas Mehtabuddin}, \bibinfo{person}{Chetan Tekur},
  \bibinfo{person}{Thomas Sohmers}, \bibinfo{person}{Kris Kang},
  \bibinfo{person}{Stephen Maresh}, {and} \bibinfo{person}{Jonathan Ross}.}
  \bibinfo{year}{2022}\natexlab{}.
\newblock \showarticletitle{A software-defined tensor streaming multiprocessor
  for large-scale machine learning}. In \bibinfo{booktitle}{\emph{Proceedings
  of the 49th Annual International Symposium on Computer Architecture}} (New
  York, New York) \emph{(\bibinfo{series}{ISCA '22})}.
  \bibinfo{publisher}{Association for Computing Machinery},
  \bibinfo{address}{New York, NY, USA}, \bibinfo{pages}{567–580}.
\newblock
\showISBNx{9781450386104}


\bibitem[Adcock et~al\mbox{.}(2026)]%
        {llama4}
\bibfield{author}{\bibinfo{person}{Aaron Adcock}, \bibinfo{person}{Aayushi
  Srivastava}, \bibinfo{person}{Abhimanyu Dubey}, \bibinfo{person}{Abhinav
  Jauhri}, \bibinfo{person}{Abhinav Pande}, \bibinfo{person}{Abhinav Pandey},
  \bibinfo{person}{Abhinav Sharma}, \bibinfo{person}{Abhishek Kadian},
  \bibinfo{person}{Abhishek Kumawat}, \bibinfo{person}{Adam Kelsey},
  {et~al\mbox{.}}} \bibinfo{year}{2026}\natexlab{}.
\newblock \showarticletitle{The Llama 4 Herd: Architecture, Training,
  Evaluation, and Deployment Notes}.
\newblock \bibinfo{journal}{\emph{arXiv preprint arXiv:2601.11659}}
  (\bibinfo{year}{2026}).
\newblock


\bibitem[Alm{\'a}si et~al\mbox{.}(2005)]%
        {collectives-on-torus}
\bibfield{author}{\bibinfo{person}{George Alm{\'a}si}, \bibinfo{person}{Philip
  Heidelberger}, \bibinfo{person}{Charles~J Archer}, \bibinfo{person}{Xavier
  Martorell}, \bibinfo{person}{C~Chris Erway}, \bibinfo{person}{Jos{\'e}~E
  Moreira}, \bibinfo{person}{Burkhard Steinmacher-Burow}, {and}
  \bibinfo{person}{Yili Zheng}.} \bibinfo{year}{2005}\natexlab{}.
\newblock \showarticletitle{Optimization of MPI collective communication on
  BlueGene/L systems}. In \bibinfo{booktitle}{\emph{Proceedings of the 19th
  annual international conference on Supercomputing}}.
  \bibinfo{pages}{253--262}.
\newblock


\bibitem[{Amazon Web Services}(2023)]%
        {inferentia2}
\bibfield{author}{\bibinfo{person}{{Amazon Web Services}}.}
  \bibinfo{year}{2023}\natexlab{}.
\newblock \bibinfo{title}{{Amazon EC2 Inf2} Architecture}.
\newblock
  \bibinfo{howpublished}{\url{https://awsdocs-neuron.readthedocs-hosted.com/en/latest/about-neuron/arch/neuron-hardware/inf2-arch.html}}.
\newblock


\bibitem[Araujo et~al\mbox{.}(2026)]%
        {MRC-network}
\bibfield{author}{\bibinfo{person}{Joao Araujo}, \bibinfo{person}{Alex Chow},
  \bibinfo{person}{Mark Handley}, \bibinfo{person}{Ryder Lewis},
  \bibinfo{person}{Christoph Paasch}, \bibinfo{person}{Jitendra Padhye},
  \bibinfo{person}{Michael Papamichael}, \bibinfo{person}{Greg Steinbrecher},
  \bibinfo{person}{Amin Tootoonchian}, \bibinfo{person}{Lihua Yuan},
  \bibinfo{person}{S. Anantharamu}, \bibinfo{person}{Abhishek Dosi},
  \bibinfo{person}{Mohit Garg}, \bibinfo{person}{Mahdieh Ghazi},
  \bibinfo{person}{Torsten Hoefler}, \bibinfo{person}{Deepal Jayasinghe},
  \bibinfo{person}{Jithin Jose}, \bibinfo{person}{Abdul Kabbani},
  \bibinfo{person}{Guohan Lu}, \bibinfo{person}{Yang Wang}, \bibinfo{person}{K.
  Doddapaneni}, \bibinfo{person}{Murali Garimella}, \bibinfo{person}{Vipin
  Jain}, \bibinfo{person}{Yanfang Le}, \bibinfo{person}{H. Nagulapalli},
  \bibinfo{person}{S. Narayanan}, \bibinfo{person}{Rong Pan},
  \bibinfo{person}{Rathina Sabesan}, \bibinfo{person}{Raghava Sivaramu},
  \bibinfo{person}{Rip Sohan}, \bibinfo{person}{Eric Davis},
  \bibinfo{person}{Dragos Dumitrescu}, \bibinfo{person}{Mohan Kalkunte},
  \bibinfo{person}{Bhaswar Mitra}, \bibinfo{person}{Guglielmo Morandin},
  \bibinfo{person}{Adrian Popa}, \bibinfo{person}{Costin Raiciu},
  \bibinfo{person}{Eric Spada}, \bibinfo{person}{John Spillane},
  \bibinfo{person}{Niranjan Vaidya}, \bibinfo{person}{Aviv Barnea},
  \bibinfo{person}{Idan Burstein}, \bibinfo{person}{Elazar Cohen},
  \bibinfo{person}{Yamin Friedman}, \bibinfo{person}{Noam Katz},
  \bibinfo{person}{Masoud Moshref}, \bibinfo{person}{Yuval Shpigelman},
  \bibinfo{person}{Shahaf Shuler}, \bibinfo{person}{Shy Shyman}, {and}
  \bibinfo{person}{Sayantan Sur}.} \bibinfo{year}{2026}\natexlab{}.
\newblock \showarticletitle{Resilient AI Supercomputer Networking using MRC and
  SRv6}.
\newblock \bibinfo{journal}{\emph{arXiv preprint arXiv:2605.04333}}
  (\bibinfo{year}{2026}).
\newblock


\bibitem[Baltz and Srivastav(2004)]%
        {multicast-congestion}
\bibfield{author}{\bibinfo{person}{Andreas Baltz} {and} \bibinfo{person}{Anand
  Srivastav}.} \bibinfo{year}{2004}\natexlab{}.
\newblock \showarticletitle{Fast Approximation of Minimum Multicast
  Congestion---Implementation versus Theory}.
\newblock \bibinfo{journal}{\emph{RAIRO Operations Research}}
  \bibinfo{volume}{38}, \bibinfo{number}{4} (\bibinfo{year}{2004}),
  \bibinfo{pages}{319--344}.
\newblock
\urldef\tempurl%
\url{https://doi.org/10.1051/ro:2004028}
\showDOI{\tempurl}


\bibitem[Basu et~al\mbox{.}(2024)]%
        {efficient-alltoall-in-direct-connect}
\bibfield{author}{\bibinfo{person}{Prithwish Basu}, \bibinfo{person}{Liangyu
  Zhao}, \bibinfo{person}{Jason Fantl}, \bibinfo{person}{Siddharth Pal},
  \bibinfo{person}{Arvind Krishnamurthy}, {and} \bibinfo{person}{Joud Khoury}.}
  \bibinfo{year}{2024}\natexlab{}.
\newblock \showarticletitle{Efficient all-to-all Collective Communication
  Schedules for Direct-connect Topologies}. In
  \bibinfo{booktitle}{\emph{Proceedings of the 33rd International Symposium on
  High-Performance Parallel and Distributed Computing}} (Pisa, Italy)
  \emph{(\bibinfo{series}{HPDC '24})}. \bibinfo{publisher}{Association for
  Computing Machinery}, \bibinfo{address}{New York, NY, USA},
  \bibinfo{pages}{28–41}.
\newblock
\showISBNx{9798400704130}


\bibitem[Benson et~al\mbox{.}(2011)]%
        {micro-te}
\bibfield{author}{\bibinfo{person}{Theophilus Benson}, \bibinfo{person}{Ashok
  Anand}, \bibinfo{person}{Aditya Akella}, {and} \bibinfo{person}{Ming Zhang}.}
  \bibinfo{year}{2011}\natexlab{}.
\newblock \showarticletitle{MicroTE: Fine grained traffic engineering for data
  centers}. In \bibinfo{booktitle}{\emph{Proceedings of the seventh conference
  on emerging networking experiments and technologies}}.
  \bibinfo{pages}{1--12}.
\newblock


\bibitem[Bernardi et~al\mbox{.}(2026)]%
        {amazon-expander}
\bibfield{author}{\bibinfo{person}{Giacomo Bernardi}, \bibinfo{person}{Ratul
  Mahajan}, \bibinfo{person}{C Seshadhri}, \bibinfo{person}{Enrico Carlesso},
  \bibinfo{person}{Chinchu~Merine Joseph}, \bibinfo{person}{Saurabh Kumar},
  \bibinfo{person}{Pavan Manikonda}, \bibinfo{person}{Luiza Popa},
  \bibinfo{person}{Randy Ram}, \bibinfo{person}{Steven Robinson},
  {et~al\mbox{.}}} \bibinfo{year}{2026}\natexlab{}.
\newblock \showarticletitle{RNG: Flat Datacenter Networks at Scale}.
\newblock \bibinfo{journal}{\emph{arXiv preprint arXiv:2604.15261}}
  (\bibinfo{year}{2026}).
\newblock


\bibitem[Boivie et~al\mbox{.}(2007)]%
        {xcast}
\bibfield{author}{\bibinfo{person}{Rick Boivie}, \bibinfo{person}{Nancy
  Feldman}, \bibinfo{person}{Yuji Imai}, \bibinfo{person}{Wim Livens}, {and}
  \bibinfo{person}{Dirk Ooms}.} \bibinfo{year}{2007}\natexlab{}.
\newblock \bibinfo{title}{Explicit Multicast ({Xcast}) Concepts and Options}.
\newblock \bibinfo{howpublished}{IETF RFC 5058}.
\newblock


\bibitem[Cai et~al\mbox{.}(2021)]%
        {dgx-collectives}
\bibfield{author}{\bibinfo{person}{Zixian Cai}, \bibinfo{person}{Zhengyang
  Liu}, \bibinfo{person}{Saeed Maleki}, \bibinfo{person}{Madanlal Musuvathi},
  \bibinfo{person}{Todd Mytkowicz}, \bibinfo{person}{Jacob Nelson}, {and}
  \bibinfo{person}{Olli Saarikivi}.} \bibinfo{year}{2021}\natexlab{}.
\newblock \showarticletitle{Synthesizing optimal collective algorithms}. In
  \bibinfo{booktitle}{\emph{Proceedings of the 26th ACM SIGPLAN Symposium on
  Principles and Practice of Parallel Programming}}. \bibinfo{pages}{62--75}.
\newblock


\bibitem[Chen et~al\mbox{.}(2000)]%
        {multicast-packing}
\bibfield{author}{\bibinfo{person}{Shiwen Chen}, \bibinfo{person}{Oktay
  Gunluk}, {and} \bibinfo{person}{Bulent Yener}.}
  \bibinfo{year}{2000}\natexlab{}.
\newblock \showarticletitle{The Multicast Packing Problem}.
\newblock \bibinfo{journal}{\emph{IEEE/ACM Transactions on Networking}}
  \bibinfo{volume}{8}, \bibinfo{number}{3} (\bibinfo{year}{2000}),
  \bibinfo{pages}{311--318}.
\newblock
\urldef\tempurl%
\url{https://doi.org/10.1109/90.851977}
\showDOI{\tempurl}


\bibitem[Chiang et~al\mbox{.}(2018)]%
        {online-multicast-te}
\bibfield{author}{\bibinfo{person}{Sheng-Hao Chiang},
  \bibinfo{person}{Jian-Jhih Kuo}, \bibinfo{person}{Shan-Hsiang Shen},
  \bibinfo{person}{De-Nian Yang}, {and} \bibinfo{person}{Wen-Tsuen Chen}.}
  \bibinfo{year}{2018}\natexlab{}.
\newblock \showarticletitle{Online multicast traffic engineering for
  software-defined networks}. In \bibinfo{booktitle}{\emph{IEEE INFOCOM
  2018-IEEE Conference on Computer Communications}}. IEEE,
  \bibinfo{pages}{414--422}.
\newblock


\bibitem[Cong et~al\mbox{.}(2024)]%
        {moe-load-stability}
\bibfield{author}{\bibinfo{person}{Peizhuang Cong}, \bibinfo{person}{Aomufei
  Yuan}, \bibinfo{person}{Shimao Chen}, \bibinfo{person}{Yuxuan Tian},
  \bibinfo{person}{Bowen Ye}, {and} \bibinfo{person}{Tong Yang}.}
  \bibinfo{year}{2024}\natexlab{}.
\newblock \showarticletitle{Prediction Is All {MoE} Needs: Expert Load
  Distribution Goes from Fluctuating to Stabilizing}.
\newblock \bibinfo{journal}{\emph{arXiv preprint arXiv:2404.16914}}
  (\bibinfo{year}{2024}).
\newblock


\bibitem[Corporation(2025)]%
        {nccl-allreduce-bench}
\bibfield{author}{\bibinfo{person}{NVIDIA Corporation}.}
  \bibinfo{year}{2025}\natexlab{}.
\newblock \bibinfo{title}{Validate the cluster level NCCL test with 4 nodes and
  32 GPUs}.
\newblock
  \bibinfo{howpublished}{\url{https://docs.nvidia.com/dgx-basepod/deployment-guide-dgx-basepod/latest/mn-nccl.html}}.
\newblock
\urldef\tempurl%
\url{https://docs.nvidia.com/dgx-basepod/deployment-guide-dgx-basepod/latest/mn-nccl.html}
\showURL{%
\tempurl}


\bibitem[{DeepSeek-AI}(2024)]%
        {deepseekv3}
\bibfield{author}{\bibinfo{person}{{DeepSeek-AI}}.}
  \bibinfo{year}{2024}\natexlab{}.
\newblock \showarticletitle{{DeepSeek-V3} Technical Report}.
\newblock \bibinfo{journal}{\emph{arXiv preprint arXiv:2412.19437}}
  (\bibinfo{year}{2024}).
\newblock


\bibitem[Ding et~al\mbox{.}(2025)]%
        {photonic-rails}
\bibfield{author}{\bibinfo{person}{Eric Ding}, \bibinfo{person}{Chuhan Ouyang},
  {and} \bibinfo{person}{Rachee Singh}.} \bibinfo{year}{2025}\natexlab{}.
\newblock \showarticletitle{Photonic rails in ML datacenters}. In
  \bibinfo{booktitle}{\emph{Proceedings of the 24th ACM Workshop on Hot Topics
  in Networks}}. \bibinfo{pages}{149--159}.
\newblock


\bibitem[Fedus et~al\mbox{.}(2022)]%
        {switch-transformer}
\bibfield{author}{\bibinfo{person}{William Fedus}, \bibinfo{person}{Barret
  Zoph}, {and} \bibinfo{person}{Noam Shazeer}.}
  \bibinfo{year}{2022}\natexlab{}.
\newblock \showarticletitle{Switch Transformers: Scaling to Trillion Parameter
  Models with Simple and Efficient Sparsity}.
\newblock \bibinfo{journal}{\emph{Journal of Machine Learning Research}}
  \bibinfo{volume}{23}, \bibinfo{number}{120} (\bibinfo{year}{2022}),
  \bibinfo{pages}{1--39}.
\newblock
\urldef\tempurl%
\url{https://www.jmlr.org/papers/v23/21-0998.html}
\showURL{%
\tempurl}


\bibitem[Fenner et~al\mbox{.}(2016)]%
        {pim-sm}
\bibfield{author}{\bibinfo{person}{Bill Fenner}, \bibinfo{person}{Mark~J.
  Handley}, \bibinfo{person}{Hugh Holbrook}, \bibinfo{person}{Isidor Kouvelas},
  \bibinfo{person}{Rishabh Parekh}, \bibinfo{person}{Zhaohui~(Jeffrey) Zhang},
  {and} \bibinfo{person}{Lianshu Zheng}.} \bibinfo{year}{2016}\natexlab{}.
\newblock \bibinfo{title}{{Protocol Independent Multicast - Sparse Mode
  (PIM-SM): Protocol Specification (Revised)}}.
\newblock \bibinfo{howpublished}{RFC 7761}.
\newblock


\bibitem[{Google Cloud}(2026)]%
        {tpu8i}
\bibfield{author}{\bibinfo{person}{{Google Cloud}}.}
  \bibinfo{year}{2026}\natexlab{}.
\newblock \bibinfo{title}{{Google} {TPU} 8i ({Boardfly}) architecture
  overview}.
\newblock
  \bibinfo{howpublished}{\url{https://cloud.google.com/blog/products/compute/tpu-8t-and-tpu-8i-technical-deep-dive}}.
\newblock


\bibitem[Hong et~al\mbox{.}(2013)]%
        {swan}
\bibfield{author}{\bibinfo{person}{Chi-Yao Hong}, \bibinfo{person}{Srikanth
  Kandula}, \bibinfo{person}{Ratul Mahajan}, \bibinfo{person}{Ming Zhang},
  \bibinfo{person}{Vijay Gill}, \bibinfo{person}{Mohan Nanduri}, {and}
  \bibinfo{person}{Roger Wattenhofer}.} \bibinfo{year}{2013}\natexlab{}.
\newblock \showarticletitle{Achieving high utilization with software-driven
  WAN}. In \bibinfo{booktitle}{\emph{Proceedings of the ACM SIGCOMM 2013}}.
  \bibinfo{pages}{15--26}.
\newblock


\bibitem[Hu et~al\mbox{.}(2025)]%
        {demystifying-nccl}
\bibfield{author}{\bibinfo{person}{Zhiyi Hu}, \bibinfo{person}{Siyuan Shen},
  \bibinfo{person}{Tommaso Bonato}, \bibinfo{person}{Sylvain Jeaugey},
  \bibinfo{person}{Cedell Alexander}, \bibinfo{person}{Eric Spada},
  \bibinfo{person}{James Dinan}, \bibinfo{person}{Jeff Hammond}, {and}
  \bibinfo{person}{Torsten Hoefler}.} \bibinfo{year}{2025}\natexlab{}.
\newblock \showarticletitle{Demystifying NCCL: An in-depth analysis of GPU
  communication protocols and algorithms}. In \bibinfo{booktitle}{\emph{2025
  IEEE Symposium on High-Performance Interconnects (HOTI)}}. IEEE,
  \bibinfo{pages}{48--59}.
\newblock


\bibitem[Jiang et~al\mbox{.}(2024)]%
        {mixtral}
\bibfield{author}{\bibinfo{person}{Albert~Q Jiang}, \bibinfo{person}{Alexandre
  Sablayrolles}, \bibinfo{person}{Antoine Roux}, \bibinfo{person}{Arthur
  Mensch}, \bibinfo{person}{Blanche Savary}, \bibinfo{person}{Chris Bamford},
  \bibinfo{person}{Devendra~Singh Chaplot}, \bibinfo{person}{Diego de~las
  Casas}, \bibinfo{person}{Emma~Bou Hanna}, \bibinfo{person}{Florian Bressand},
  {et~al\mbox{.}}} \bibinfo{year}{2024}\natexlab{}.
\newblock \showarticletitle{Mixtral of experts}.
\newblock \bibinfo{journal}{\emph{arXiv preprint arXiv:2401.04088}}
  (\bibinfo{year}{2024}).
\newblock


\bibitem[Jouppi et~al\mbox{.}(2023)]%
        {tpuv4}
\bibfield{author}{\bibinfo{person}{Norm Jouppi}, \bibinfo{person}{George
  Kurian}, \bibinfo{person}{Sheng Li}, \bibinfo{person}{Peter Ma},
  \bibinfo{person}{Rahul Nagarajan}, \bibinfo{person}{Lifeng Nai},
  \bibinfo{person}{Nishant Patil}, \bibinfo{person}{Suvinay Subramanian},
  \bibinfo{person}{Andy Swing}, \bibinfo{person}{Brian Towles},
  \bibinfo{person}{Clifford Young}, \bibinfo{person}{Xiang Zhou},
  \bibinfo{person}{Zongwei Zhou}, {and} \bibinfo{person}{David~A Patterson}.}
  \bibinfo{year}{2023}\natexlab{}.
\newblock \showarticletitle{TPU v4: An Optically Reconfigurable Supercomputer
  for Machine Learning with Hardware Support for Embeddings}. In
  \bibinfo{booktitle}{\emph{Proceedings of the 50th Annual International
  Symposium on Computer Architecture}} (Orlando, FL, USA)
  \emph{(\bibinfo{series}{ISCA '23})}. \bibinfo{publisher}{Association for
  Computing Machinery}, \bibinfo{address}{New York, NY, USA}, Article
  \bibinfo{articleno}{82}, \bibinfo{numpages}{14}~pages.
\newblock
\showISBNx{9798400700958}


\bibitem[Li et~al\mbox{.}(2011)]%
        {steiner-tree-nphard}
\bibfield{author}{\bibinfo{person}{Dan Li}, \bibinfo{person}{Yuanjie Li},
  \bibinfo{person}{Jianping Wu}, \bibinfo{person}{Sen Su}, {and}
  \bibinfo{person}{Jiangwei Yu}.} \bibinfo{year}{2011}\natexlab{}.
\newblock \showarticletitle{ESM: Efficient and scalable data center multicast
  routing}.
\newblock \bibinfo{journal}{\emph{IEEE/ACM Transactions on Networking}}
  \bibinfo{volume}{20}, \bibinfo{number}{3} (\bibinfo{year}{2011}),
  \bibinfo{pages}{944--955}.
\newblock


\bibitem[Li et~al\mbox{.}(2023)]%
        {lina}
\bibfield{author}{\bibinfo{person}{Jiamin Li}, \bibinfo{person}{Yimin Jiang},
  \bibinfo{person}{Yibo Zhu}, \bibinfo{person}{Cong Wang}, {and}
  \bibinfo{person}{Hong Xu}.} \bibinfo{year}{2023}\natexlab{}.
\newblock \showarticletitle{Accelerating Distributed {MoE} Training and
  Inference with {Lina}}. In \bibinfo{booktitle}{\emph{2023 USENIX Annual
  Technical Conference (USENIX ATC 23)}}. \bibinfo{publisher}{USENIX
  Association}, \bibinfo{pages}{945--959}.
\newblock


\bibitem[Liao et~al\mbox{.}(2025)]%
        {mixnet}
\bibfield{author}{\bibinfo{person}{Xudong Liao}, \bibinfo{person}{Yijun Sun},
  \bibinfo{person}{Han Tian}, \bibinfo{person}{Xinchen Wan},
  \bibinfo{person}{Yilun Jin}, \bibinfo{person}{Zilong Wang},
  \bibinfo{person}{Zhenghang Ren}, \bibinfo{person}{Xinyang Huang},
  \bibinfo{person}{Wenxue Li}, \bibinfo{person}{Kin~Fai Tse}, {et~al\mbox{.}}}
  \bibinfo{year}{2025}\natexlab{}.
\newblock \showarticletitle{Mixnet: A runtime reconfigurable optical-electrical
  fabric for distributed mixture-of-experts training}. In
  \bibinfo{booktitle}{\emph{Proceedings of the ACM SIGCOMM 2025 Conference}}.
  \bibinfo{pages}{554--574}.
\newblock


\bibitem[{Moonshot AI}(2026)]%
        {kimik3}
\bibfield{author}{\bibinfo{person}{{Moonshot AI}}.}
  \bibinfo{year}{2026}\natexlab{}.
\newblock \bibinfo{title}{{Kimi K3}: Open Frontier Intelligence}.
\newblock \bibinfo{howpublished}{\url{https://www.kimi.com/blog/kimi-k3}}.
\newblock


\bibitem[Pope et~al\mbox{.}(2023)]%
        {scaling-transformer-inference}
\bibfield{author}{\bibinfo{person}{Reiner Pope}, \bibinfo{person}{Sholto
  Douglas}, \bibinfo{person}{Aakanksha Chowdhery}, \bibinfo{person}{Jacob
  Devlin}, \bibinfo{person}{James Bradbury}, \bibinfo{person}{Anselm Levskaya},
  \bibinfo{person}{Jonathan Heek}, \bibinfo{person}{Kefan Xiao},
  \bibinfo{person}{Shivani Agrawal}, {and} \bibinfo{person}{Jeff Dean}.}
  \bibinfo{year}{2023}\natexlab{}.
\newblock \showarticletitle{Efficiently Scaling Transformer Inference}. In
  \bibinfo{booktitle}{\emph{Proceedings of Machine Learning and Systems
  (MLSys)}}.
\newblock


\bibitem[{Qwen Team}(2025)]%
        {qwen3}
\bibfield{author}{\bibinfo{person}{{Qwen Team}}.}
  \bibinfo{year}{2025}\natexlab{}.
\newblock \showarticletitle{Qwen3 Technical Report}.
\newblock \bibinfo{journal}{\emph{arXiv preprint arXiv:2505.09388}}
  (\bibinfo{year}{2025}).
\newblock


\bibitem[{Qwen Team}(2026)]%
        {qwen35}
\bibfield{author}{\bibinfo{person}{{Qwen Team}}.}
  \bibinfo{year}{2026}\natexlab{}.
\newblock \bibinfo{title}{{Qwen3.5-397B-A17B} model card}.
\newblock
  \bibinfo{howpublished}{\url{https://huggingface.co/Qwen/Qwen3.5-397B-A17B}}.
\newblock


\bibitem[Sridharan et~al\mbox{.}(2023)]%
        {chakra}
\bibfield{author}{\bibinfo{person}{Srinivas Sridharan},
  \bibinfo{person}{Taekyung Heo}, \bibinfo{person}{Louis Feng},
  \bibinfo{person}{Zhaodong Wang}, \bibinfo{person}{Matt Bergeron},
  \bibinfo{person}{Wenyin Fu}, \bibinfo{person}{Shengbao Zheng},
  \bibinfo{person}{Brian Coutinho}, \bibinfo{person}{Saeed Rashidi},
  \bibinfo{person}{Changhai Man}, {and} \bibinfo{person}{Tushar Krishna}.}
  \bibinfo{year}{2023}\natexlab{}.
\newblock \showarticletitle{Chakra: Advancing Performance Benchmarking and
  Co-design using Standardized Execution Traces}.
\newblock \bibinfo{journal}{\emph{arXiv preprint arXiv:2305.14516}}
  (\bibinfo{year}{2023}).
\newblock


\bibitem[Suh and Valamanchili(1998)]%
        {a2a-on-tori}
\bibfield{author}{\bibinfo{person}{Young-Joo Suh} {and} \bibinfo{person}{S
  Valamanchili}.} \bibinfo{year}{1998}\natexlab{}.
\newblock \showarticletitle{All-to-all communication with minimum start-up
  costs in 2D/3D tori and meshes}.
\newblock \bibinfo{journal}{\emph{IEEE Transactions on Parallel and Distributed
  Systems}} \bibinfo{volume}{9}, \bibinfo{number}{5} (\bibinfo{year}{1998}),
  \bibinfo{pages}{442--458}.
\newblock


\bibitem[Tang et~al\mbox{.}(2025)]%
        {pangu}
\bibfield{author}{\bibinfo{person}{Yehui Tang}, \bibinfo{person}{Yichun Yin},
  \bibinfo{person}{Yaoyuan Wang}, \bibinfo{person}{Hang Zhou},
  \bibinfo{person}{Yu Pan}, \bibinfo{person}{Wei Guo}, \bibinfo{person}{Ziyang
  Zhang}, \bibinfo{person}{Miao Rang}, \bibinfo{person}{Fangcheng Liu},
  \bibinfo{person}{Naifu Zhang}, \bibinfo{person}{Binghan Li},
  \bibinfo{person}{Yonghan Dong}, \bibinfo{person}{Xiaojun Meng},
  \bibinfo{person}{Yasheng Wang}, \bibinfo{person}{Dong Li},
  \bibinfo{person}{Yin Li}, \bibinfo{person}{Dandan Tu}, \bibinfo{person}{Can
  Chen}, \bibinfo{person}{Youliang Yan}, \bibinfo{person}{Fisher Yu},
  \bibinfo{person}{Ruiming Tang}, \bibinfo{person}{Yunhe Wang},
  \bibinfo{person}{Botian Huang}, \bibinfo{person}{Bo Wang},
  \bibinfo{person}{Boxiao Liu}, \bibinfo{person}{Changzheng Zhang},
  \bibinfo{person}{Da Kuang}, \bibinfo{person}{Fei Liu}, \bibinfo{person}{Gang
  Huang}, \bibinfo{person}{Jiansheng Wei}, \bibinfo{person}{Jiarui Qin},
  \bibinfo{person}{Jie Ran}, \bibinfo{person}{Jinpeng Li}, \bibinfo{person}{Jun
  Zhao}, \bibinfo{person}{Liang Dai}, \bibinfo{person}{Lin Li},
  \bibinfo{person}{Liqun Deng}, \bibinfo{person}{Peifeng Qin},
  \bibinfo{person}{Pengyuan Zeng}, \bibinfo{person}{Qiang Gu},
  \bibinfo{person}{Shaohua Tang}, \bibinfo{person}{Shengjun Cheng},
  \bibinfo{person}{Tao Gao}, \bibinfo{person}{Tao Yu}, \bibinfo{person}{Tianshu
  Li}, \bibinfo{person}{Tianyu Bi}, \bibinfo{person}{Wei He},
  \bibinfo{person}{Weikai Mao}, \bibinfo{person}{Wenyong Huang},
  \bibinfo{person}{Wulong Liu}, \bibinfo{person}{Xiabing Li},
  \bibinfo{person}{Xianzhi Yu}, \bibinfo{person}{Xueyu Wu}, \bibinfo{person}{Xu
  He}, \bibinfo{person}{Yangkai Du}, \bibinfo{person}{Yan Xu},
  \bibinfo{person}{Ye Tian}, \bibinfo{person}{Yimeng Wu},
  \bibinfo{person}{Yongbing Huang}, \bibinfo{person}{Yong Tian},
  \bibinfo{person}{Yong Zhu}, \bibinfo{person}{Yue Li}, \bibinfo{person}{Yufei
  Wang}, \bibinfo{person}{Yuhang Gai}, \bibinfo{person}{Yujun Li},
  \bibinfo{person}{Yu Luo}, \bibinfo{person}{Yunsheng Ni},
  \bibinfo{person}{Yusen Sun}, \bibinfo{person}{Zelin Chen},
  \bibinfo{person}{Zhe Liu}, \bibinfo{person}{Zhicheng Liu},
  \bibinfo{person}{Zhipeng Tu}, \bibinfo{person}{Zilin Ding}, {and}
  \bibinfo{person}{Zongyuan Zhan}.} \bibinfo{year}{2025}\natexlab{}.
\newblock \showarticletitle{Pangu ultra MoE: How to train your big MoE on
  ASCEND NPUS}.
\newblock \bibinfo{journal}{\emph{arXiv preprint arXiv:2505.04519}}
  (\bibinfo{year}{2025}).
\newblock


\bibitem[Vahdat(2026)]%
        {tpu8-bw}
\bibfield{author}{\bibinfo{person}{Amin Vahdat}.}
  \bibinfo{year}{2026}\natexlab{}.
\newblock \bibinfo{title}{Our eighth generation TPUs: two chips for the agentic
  era}.
\newblock
  \bibinfo{howpublished}{\url{https://blog.google/innovation-and-ai/infrastructure-and-cloud/google-cloud/eighth-generation-tpu-agentic-era/}}.
\newblock
\urldef\tempurl%
\url{https://blog.google/innovation-and-ai/infrastructure-and-cloud/google-cloud/eighth-generation-tpu-agentic-era/}
\showURL{%
\tempurl}


\bibitem[Wang et~al\mbox{.}(2024)]%
        {wang2024auxiliary}
\bibfield{author}{\bibinfo{person}{Lean Wang}, \bibinfo{person}{Huazuo Gao},
  \bibinfo{person}{Chenggang Zhao}, \bibinfo{person}{Xu Sun}, {and}
  \bibinfo{person}{Damai Dai}.} \bibinfo{year}{2024}\natexlab{}.
\newblock \showarticletitle{Auxiliary-loss-free load balancing strategy for
  mixture-of-experts}.
\newblock \bibinfo{journal}{\emph{arXiv preprint arXiv:2408.15664}}
  (\bibinfo{year}{2024}).
\newblock


\bibitem[Wang et~al\mbox{.}(2023)]%
        {topoopt}
\bibfield{author}{\bibinfo{person}{Weiyang Wang}, \bibinfo{person}{Moein
  Khazraee}, \bibinfo{person}{Zhizhen Zhong}, \bibinfo{person}{Manya Ghobadi},
  \bibinfo{person}{Zhihao Jia}, \bibinfo{person}{Dheevatsa Mudigere},
  \bibinfo{person}{Ying Zhang}, {and} \bibinfo{person}{Anthony Kewitsch}.}
  \bibinfo{year}{2023}\natexlab{}.
\newblock \showarticletitle{$\{$TopoOpt$\}$: Co-optimizing network topology and
  parallelization strategy for distributed training jobs}. In
  \bibinfo{booktitle}{\emph{20th USENIX Symposium on Networked Systems Design
  and Implementation (NSDI 23)}}. \bibinfo{pages}{739--767}.
\newblock


\bibitem[Wijnands et~al\mbox{.}(2017)]%
        {bier}
\bibfield{author}{\bibinfo{person}{IJsbrand Wijnands}, \bibinfo{person}{Eric~C.
  Rosen}, \bibinfo{person}{Andrew Dolganow}, \bibinfo{person}{Tony Przygienda},
  {and} \bibinfo{person}{Sam Aldrin}.} \bibinfo{year}{2017}\natexlab{}.
\newblock \bibinfo{title}{Multicast Using Bit Index Explicit Replication
  ({BIER})}.
\newblock \bibinfo{howpublished}{IETF RFC 8279}.
\newblock


\bibitem[Won et~al\mbox{.}(2023)]%
        {astrasim2}
\bibfield{author}{\bibinfo{person}{William Won}, \bibinfo{person}{Taekyung
  Heo}, \bibinfo{person}{Saeed Rashidi}, \bibinfo{person}{Srinivas Sridharan},
  \bibinfo{person}{Sudarshan Srinivasan}, {and} \bibinfo{person}{Tushar
  Krishna}.} \bibinfo{year}{2023}\natexlab{}.
\newblock \showarticletitle{{ASTRA-sim2.0}: Modeling Hierarchical Networks and
  Disaggregated Systems for Large-model Training at Scale}. In
  \bibinfo{booktitle}{\emph{IEEE International Symposium on Performance
  Analysis of Systems and Software (ISPASS)}}.
\newblock


\end{thebibliography}

\end{document}